\documentstyle[12pt]{article}

\begin{document}

\topmargin 0pt
\oddsidemargin 5mm

\setcounter{page}{1}
\vspace{2cm}
\begin{center}

{\bf FINITE SIZE CORRECTIONS ON THE BOUNDARY BETWEEN
THE SPIN-GLASS AND THE FERROMAGNETIC PHASES OF DERRIDA's MODEL}\\
\vspace{5mm}
{\large Y.M. Hakobyan,D.B. Sahakyan}\\
\vspace{3mm}
{\em Yerevan Physics Institute}
Alikhanian Brothers St.2, Yerevan 375036, Armenia\\
Saakian @ vx1.YERPHI.AM\\
{\large M.R. Daj}\\
{\em Department of Physics, Sharif University of Technology}\\
P. O. Box 9161, Tehran 11365, Iran.
\end{center}

\vspace{10mm}
\centerline{{\bf Abstract}}
The boundary line between the ferromagnetic and the spin-glass
phases was investigated. Finite size corrections to the free
 energy and magnetization were calculated.

The situation coincides with the case in the information theory,
when the transmission rate equals to the capacity of channel.
\vspace{10mm}

{\bf 1. Introduction.}\\
\indent
Derrida's model [1] is connected with optimal coding, as was suggested
in [2] and proved in [3]. Different aspects of this connection were considered
in [4,5].

In [6,7] were considered finite size effects to the magnetization for the fully
connected and diluted models in the ferromagnetic phase. This is interesting
both for physics and information theory applications. It corresponds to decoding
error probability.
Rich fine structure of the ferromagnetic phase was considered in [8].
In [9] was calculated vanishing magnetization in the spin-glass phase.

It's also very interesting to calculate finite size effects on the boundary
line between the ferromagnetic and the spin-glass phases.
\newpage {\bf 2. Finite size corrections to the free energy.}\\
\vspace{1cm}
We consider Derrida's model with the Hamiltonian:
\begin{equation}
H=-\sum_{1\le i_{1}\le i_{2}\le i_{3}\cdots\le N}
\left( J_{0}\frac{N}{C_{N}^{p}}+J_{i_1\dots i_p}
\sqrt{\frac{N}{C_N^p}}\right)\,
\sigma_{i_{1}}\dots\sigma_{i_{p}}
\end{equation}
where $J_{i_{1}\dots i_{p}}$ are quenched random coupling
constants with a Gaussian distribution:
\begin{equation}
<J_{i_{1}\dots i_{p}}^2>=\frac{1}{2}
\end{equation}
The ferromagnetic phase appears (in low temperatures) [8]:
\begin{equation}
J_{0}>\sqrt{\ln{2}}
\end{equation}
We consider just the case:
\begin{equation}
J_{0}=\sqrt{\ln{2}}
\end{equation}
As argued Derrida, with high (exponential) accuracy our system (1)
is equivalent to Random Energy Model (REM) of $2^N$ energy
levels, distributed independently.\\
The first energy level has the probability distribution:
\begin{equation}
P(E)=\frac{1}{\sqrt{\pi N}}\, e^{-\frac{(E+J_{0} N B)^2}{N}}
\end{equation}
and another $2^N-1$ levels:
\begin{equation}
P(E)=\frac{1}{\sqrt{\pi N}}\, e^{-\frac{E^2}{N}}
\end{equation}
We therefore find that the free energy  of REM is given by:
\begin{equation}
<\ln{Z}>=<\ln{\sum_{\alpha=1}^{2^N}} e^{-BE_{\alpha}}>
\end{equation}
and the averaging is over the distribution of $E_{\alpha}$.

To perform the averaging in (7) we use the trick introduced by
Derrida [1]:
\begin{equation}
<\ln{Z}>=\Gamma'(1)+\int\limits_{\infty}^{\infty}u\, d[e^{-\phi(u)}]
\end{equation}
where $u=\ln{t}$; $e^{-\phi}=<e^{-t Z}>$ and $\Gamma'$ is derivative
of the gamma function.\\
It is easy to see that:
\begin{equation}
e^{-\phi(u)}=f(u+J_{0} N B) f(u)^{2^N-1}
\end{equation}
where
\begin{equation}
f(u)=\frac{1}{\sqrt{\pi }}\quad \int\limits_{-\infty }^{\infty}
dy\, e^{-y^2-e^{-\lambda y}+u}; \lambda=B\sqrt{N}
\end{equation}
Derrida has given approximate expressions for the function $f(u)$:
\begin{eqnarray}
f(u)\simeq\frac{\Gamma({\frac{2u}{\lambda^2}})}{\sqrt{\pi}{\lambda}}
e^{-\frac{u^2}{\lambda^2}};&u>0\\
f(u)\simeq1-\frac{\mid\Gamma(\frac{2u}{\lambda^2})\mid}{\sqrt{\pi}{\lambda}}
e^{-\frac{u^2}{\lambda^2}};&\frac{\lambda^2}{2}<u<0\\
f(u)\simeq1-e^{u+\frac{\lambda^2}{4}};&-\lambda^2<u<-\frac{\lambda^2}{2}
\end{eqnarray}
In the appendix we have derived the expression for $f(u)$ in the
case $1<<\frac{\mid u\mid}{\lambda}<<\lambda$.\\
To evaluate approximately (8) we write it in a suitable form:
\begin{equation}
<\ln{Z}>=u_{0}+\int\limits_{-\infty}^{\infty}f(u+u_{0})\Psi(u)\,du
\end{equation}
where $u_{0}=B J_{0} N$ and $\Psi(u)=1-f(u)^{2^N-1}$.\\
Let's denote:
                        $$
f(u+u_{0})=F'(u);\,F(u)=\int\limits_{0}^{u+u_{0}}f(x) dx
                        $$
For our purposes it's enough to know the values of $F(u)$ near
to $-u_{0}$.

To calculate the integral in (14) it is convenient to integrate
by parts as the derivation of  the function  $\Psi(u)$ behaves
like $\delta-$ function.

The thing is that both functions $F'(u)$
and $\Psi(u)$ are like step-functions at the same point (as we
are on the  boundary between the phases).
Moreover we can replace the integration
$\int\limits_{-\infty}^{\infty}$ by $\int\limits_{-u_{0}-\delta}^{-u_{0}+\delta}$
because of the behavior of  our functions f(u) and $\Psi(u)$
(where $\lambda<<\delta<<\lambda^2$),  outside the  interval
$[-u_{0}-\delta, -u_{0}+\delta]\,$  $f(u)*\Psi(u)$ vanishes
exponentially.

That is:
\begin{equation}
\int\limits_{-\infty}^{\infty}F'(u)\Psi(u)\,du=
\left.F(u)\Psi(u)\right|_{-u_0-\delta}^{-u_0+\delta}
-\int\limits_{-u_{0}-\delta}^{\-u_{0}+\delta}F(u)\Psi'(u)\,du
\end{equation}
As has been proved in the appendix:
\begin{equation}
F(u)\simeq\int\limits_{0}^{u+u_{0}}dx\{\frac{1}{\sqrt{\pi}}
\int\limits_{\frac{x}{\lambda}}^{\infty}e^{-t^2}\,dt
-\frac{C}{\sqrt{\pi}}\exp[-u^2/\lambda^2]\}
\end{equation}
where $C$ is Euler constant.
As  the function $\Psi(u)$  is very close to  a step function:
it is very   near one for $u>-u_{0}$ and near zero for
$u<-u_{0}$, so
\begin{equation}
F(-u_{0}+\delta)\Psi(-u_{0}+\delta)\simeq F(-u_{0}+\delta)=
\simeq\frac{\lambda}{\sqrt\pi}\int\limits_{0}^{\infty}dx\int\limits_{x}^{\infty}e^{-t^2}\,dt
-C/2
\end{equation}\\
			$
F(-u_{0}-\delta)\Psi(-u_{0}-\delta)\simeq0\,$ because
$\,\Psi(-u_{0}-\delta)\sim
e^{-\frac{\delta^2}{\lambda^2}}\rightarrow 0$\\
The integral in (15) we  denote by:
\begin{equation}
I=-\int\limits_{-u_{0}-\delta}^{-u_{0}+\delta}F(u)\,d\Psi(u)=
\int\limits_{-u_{0}-\delta}^{-u_{0}+\delta}\,F(u)df(u)^{2^N}
\end{equation}\\
and after expansion the function F(u) near the point $u=-u_{0}$\\
\begin{equation}
I\simeq-\int\limits_{-u_{0}-\delta}^{-u_{0}+\delta}
[F(-u_{0})+F'(-u_{0})(u+u_{0})+F''(-u_{0})\frac{(u+u_{0})^2}{2}]\,df(u)^{M}
\end{equation}

We'll need $u$  expressed in terms of $\phi$.\\
Using Derrida's tricks
\begin{eqnarray}
f^M=e^{-\phi}=e^{-\frac{A}{\sqrt{N}}-\frac{u^2}{\lambda^2}+N\ln{2}};
            \mbox{where} A\sim 1;\nonumber\\
\ln{\phi}=-\frac{u^2}{\lambda^2}+N\ln2-\frac{1}{2}\ln{N}+\ln A;\nonumber\\
\frac{u^2}{\lambda^2}=N\ln{2}-\frac{1}{2}\ln{N}+\ln{A}-\ln{\phi}\nonumber
\end{eqnarray}\\
It is easy to get that:
\begin{equation}
u\simeq -BN\sqrt{\ln 2}+\frac{B\ln N}{4\sqrt{\ln 2}}
+\frac{B\ln\phi}{2\sqrt{\ln 2}}
\end{equation}
Using this we obtain:
\begin{eqnarray}
I=-F(-u_{0})+F'(-u_{0})\int\limits^{-u_{0}+\delta}_{-u_{0}-\delta}(u+u_{0})\,df^M+
\frac{F''(-u_{0})}{2}\int\limits^{-u_{0}+\delta}_{-u_{0}-\delta}(u+u_{0})^2\,df^{2^N}=\\
=0+f(0)I_1+f'(0)I_2\nonumber
\end{eqnarray}
where by $I_1,I_2$ are denoted the integrals, given above.\\
Using (19) we find that:
\begin{equation}
I_{1}\simeq-\frac{B\ln N}{4\sqrt{\ln 2}}
\end{equation}
In the appendix we derived the expressions for f(0) and f'(0).\\
After  all this tricks and conserving the major elements we obtain:
\begin{eqnarray}
<\ln Z>=-u_{0}-\frac{1}{4}\frac{B}{B_{c}}\ln N+\frac{B\sqrt{N}}{\sqrt\pi}\int\limits_{0}^{\infty}dx\int\limits_{x}^{\infty}e^{-t^2}\,dt
-C/2; \nonumber\\
B_{c}=2\sqrt{\ln2};\\
u_{0}=J_{0}BN\nonumber
\end{eqnarray}
We have got a strange result: the finite size effects on the SG-FM boundary are
of the order of $\sqrt{N}$.
\newpage
{\bf 3. Calculation of the magnetization.}\\

The magnetization for particularly values of $E_{1}\dots E_{M}$
equals:
\begin{equation}
m=\left<\frac{e^{-BE_{1}}-e^{-BE_2}}
{\sum_{\alpha=1}^{M}e^{-BE_{\alpha}}}\right>;\quad (M=2^N)
\end{equation}
With the  accuracy $\sim 2^{-N}$ we neglect by second term in the
nominator and transform (23) to the form:
\begin{eqnarray}
m=\int P(E_{1}\dots E_{M})
e^{-BE_{1}}\int\limits_{0}^{\infty}e^{-t\sum_{\alpha}e^{-BE_{\alpha}}}=
-\int\limits_{-\infty}^{\infty}\,du\, f'(u+u_{0})f^{M-1}(u)=\nonumber\\
=1-\int\limits_{-\infty}^{\infty}\,du\quad  f'(u+u_{0})[f^{M-1}(u)-1]=
1+\int\limits_{-\infty}^{\infty} f(u+u_{0})\,df^{M-1}=\\
=1+\int\limits_{-\infty}^{\infty} f(u+u_{0})\,de^{-\phi}\nonumber
\end{eqnarray}
Let's expand the  function $f(u+u_{0})$ near the point $u=-u_{0}$:
\begin{equation}
f(u+u_{0})=f(0)+f'(0)(u+u_{0})+f'(0)\frac{(u+u_{0})^2}{2}
\end{equation}
Using formula (19), which gives $u$ expressed by $\phi$, we
obtain:
\begin{equation}
m=\frac{1}{2}+\frac{1}{\sqrt{\pi N}B}C-
\frac{1}{\sqrt{\pi N} 2B_{c}}\ln N
\end{equation}

where C is Euler's constant (see Appendix A).
In the limit $B\rightarrow \infty$ the second term disappears.\\
This formula is the main result of our work.
\begin{center}
{\bf Acknowledgements}
\end{center}
D. B. Sahakyan
thanks ISF support (Grant MVMOOO) and German Ministry of Research
and Technology grant 211-5231 for partial support.
\newpage

\setcounter{equation}{0}
\appendix{{\bf Appendix A}}
\renewcommand{\theequation}{A.\arabic{equation}}

\indent
In this appendix we calculate the values of $f(0)$ and $f'(0)$,where
\begin{equation}
f(u)=\frac{1}{\sqrt{\pi }}\quad \int\limits^{\infty }_{-\infty
}e^{-x^2-e^{\lambda x+u}}\, dx
\end{equation}
By changing the variables $e^{\lambda x+u}=y $, we obtain
\begin{equation}
f(u)=\frac{1}{\sqrt{\pi}\lambda}\quad e^{-u^2/\lambda ^2}
\int\limits_{0}^{\infty} e^{-\frac{(\ln x)^2}{\lambda ^2}
+ (\frac{2u}{\lambda ^2}-1)\ln x-x} \,dx
\end{equation}
Then split it like this:
\begin{equation}
f(u)=\frac{1}{\sqrt{\pi}\lambda}\quad e^{-u^2/\lambda ^2}
\int\limits_{0}^{1} \dots+\frac{1}{\sqrt{\pi}\lambda}\quad e^{-u^2/\lambda ^2}
\int\limits_{1}^{\infty} \dots
\end{equation}
By changing the variables in the first integral (which we denote by
$J_{1}$) according to $x=y^{\lambda}$, we get:
\begin{equation}
J_{1}=\frac{1}{\sqrt{\pi}}\quad e^{-u^2/\lambda ^2}
\int\limits_{0}^{1} e^{-(\ln y)^2+(\frac{2u}{\lambda}-1)\ln
                                  y-y^{\lambda}} \,dy
\end{equation}
After the
expansion
$e^{-y^{\lambda}}=\sum^{\infty}_{k=0}\frac{(-1)^k}{k!}e^{\lambda k\ln
y}= \sum^{\infty}_{k=0}\frac{(-1)^k}{k!}y^{\lambda k}$ and replacement
$\ln y=x$, we have $$ J_{1}=\frac{1}{\sqrt{\pi}}\quad e^{-u^2/\lambda
^2}\sum_{k=0}^{\infty}\frac{(-1)^k}{k!} \int\limits_{0}^{\infty}
e^{-x^2-\frac{2u}{\lambda}x-\lambda kx} \,dx $$ For f(u) we obtain:
\begin{eqnarray}
f(u)=\frac{1}{\sqrt{\pi}}\int\limits_{\frac{u}{\lambda}}^{\infty}e^{-z^2}\,dz+
\frac{1}{\sqrt{\pi}}\quad e^{-u^2/\lambda ^2}\sum_{k=1}^{\infty}\frac{(-1)^k}{k!}
e^{(\frac{u}{\lambda}+\frac{k\lambda}{2})^2}
\int\limits_{\frac{u}{\lambda}+\frac{k\lambda}{2}}^{\infty} e^{-z^2}dz
\nonumber\\
+\frac{1}{\sqrt{\pi}\lambda}\quad e^{-u^2/\lambda ^2}
\int\limits_{1}^{\infty} e^{-\frac{(\ln x)^2}{\lambda ^2}
+ (\frac{2u}{\lambda ^2}-1)\ln x-x} \,dx
\end{eqnarray}
Now it is not difficult to check that:
\begin{eqnarray}
f(0)\simeq\frac{1}{2}+\frac{1}{\sqrt\pi\lambda}\sum^{\infty}_{k=1}
\frac{(-1)^k}{k!k}
+\frac{1}{\sqrt\pi\lambda}\int\limits_{1}^{\infty}\frac{e^{-x}}{x}
\,dx+O(\frac{1}{\lambda^2});\nonumber
f'(0)\simeq-\frac{1}{\sqrt{\pi}\lambda}+O(\frac{1}{\lambda^2})\nonumber
\end{eqnarray}
If we note that:
\begin{equation}
C=-\int\limits_{1}^{\infty}\frac{e^{-x}}{x}\,dx-\sum_{k=1}^{\infty}
\frac{(-1)^k}{k!k}
\end{equation}
then
\begin{eqnarray}
f(0)\simeq\frac{1}{2}-\frac{1}{\sqrt{\pi}\lambda}C+O(\frac{1}{\lambda^2})\\
f'(0)\simeq-\frac{1}{\sqrt{\pi}\lambda}+O(\frac{1}{\lambda^2})
\end{eqnarray}
where $C=0.577\dots$ is Euler's constant.\\ We also need the integral:
\begin{equation}
\int_0^{\infty}f(u)du\simeq\frac{B\sqrt{N}}{\sqrt\pi}\int\limits_{0}^{\infty}dx\int\limits_{x}^{\infty}e^{-t^2}\,dt
-C/2                      
\end{equation}

\end{document}